# Generación de voces artificiales infantiles en castellano con acento costarricense

Ana Lilia Álvarez Blanco[1,2], Eugenia Córdoba Warner[1,2], Marvin Coto Jiménez[1,3], Vivian Fallas López[1,2], Maribel Morales Rodríguez[1,2].

[1]TecnoinclusiónUCR, [2]Escuela de Orientación y Educación Especial, Universidad de Costa Rica, [3]Escuela de Ingeniería Eléctrica, Universidad de Costa Rica.

tecnoinclusion.eoee@ucr.ac.cr

**Resumen**

En el presente artículo se evalúa una primera experiencia de generación de voces artificiales infantiles con acento costarricense, utilizando la técnica de síntesis estadística paramétrica de voz basada en Modelos Ocultos de Markov. Se describe el proceso de grabación de las muestras de voz utilizadas para el aprendizaje de los modelos, los fundamentos de la técnica empleada y la evaluación subjetiva de los resultados a través de la percepción de un grupo de personas. Los resultados muestran que la inteligibilidad de los resultados, evaluados en palabras aisladas, se encuentran por debajo de las voces grabadas por el grupo de niños y niñas participantes. De igual manera, la detección de la edad y el género de la persona hablante se ve afectada significativamente en las voces artificiales, en relación con grabaciones de voces naturales. Estos resultados muestran las necesidades de obtener mayores cantidades de datos, además de constituirse en una referencia numérica para futuros desarrollos producto de nuevos datos o de procesos de mejora de los resultados en la misma técnica.

**Palabras clave:** Costa Rica, habla, HMM, síntesis de voz, voces infantiles.

**Abstract**

This article evaluates a first experience of generating artificial children's voices with a Costa Rican accent, using the technique of statistical parametric speech synthesis based on Hidden Markov Models. The process of recording the voice samples used for learning the models, the fundamentals of the technique used and the subjective evaluation of the results through the perception of a group of people is described. The results show that the intelligibility of the results, evaluated in isolated words, is lower than the voices recorded by the group of participating children. Similarly, the detection of the age and gender of the speaking person is significantly affected in artificial voices, relative to recordings of natural voices. These results show the need to obtain larger amounts of data, in addition to becoming a numerical reference for future developments resulting from new data or from processes to improve results in the same technique.

## 1. Introducción

Las tecnologías del habla son un conjunto de técnicas y procedimientos desarrollados para la comunicación verbal con y mediante los dispositivos tecnológicos. Además del análisis de las señales de habla para distintos fines, tales como el diagnóstico médico y los sistemas de identificación biométrica, la comunicación verbal en sus dos vías: producción de habla, escucha y comprensión, se han implementado como sistemas de generación de voz artificial y de reconocimiento automático del habla. De esta manera, el estudio de tecnologías del habla podría dividirse en primer lugar en: sistemas de reconocimiento, sistemas de síntesis, y sistemas de análisis de las señales de voz.

En cuanto a los sistemas de síntesis, que emulan el proceso de emisión de habla por parte de las personas, estos tienen como objetivo el convertir una representación en texto a lo interno de un dispositivo, en una señal acústica que sea indistinguible del habla humana (Macchi, 1998). Este proceso de conversión es necesario dado que todas las representaciones de información a lo interno de una computadora o dispositivo electrónico se encuentran en forma de representaciones numéricas, a las cuales se les puede dar una relación con letras y palabras de idiomas específicos, y de éstos es necesaria la conversión a sonidos de habla para que puedan ser emitidos en una comunicación verbal.

Las aplicaciones de la síntesis de voz, o generación de habla artificial, se encuentran en múltiples áreas, como los sistemas de asistencia para la población con discapacidad (Condado y Lobo, 2007) (Jreige, Patel, y Bunnell, 2009) (Kobayashi, Fukuda, Takagi, y Asakawa, 2009), el desarrollo de audiolibros (Evans y Reichenbach, 2012), la implementación entornos educativos y de entretenimiento (Narayanan y Alwan, 2005), (Sefara, Mokgonyane, Manamela, y Modipa, 2019) y los asistentes personales virtuales (de Barcelos, et al.,2020). En general pueden establecerse como áreas potenciales de interés cualquier aplicación que contemple interacción humano-computadora, la cual puede beneficiar a las personas usuarias al permitir la interacción utilizando la forma de comunicación más usual entre seres humanos: el habla.

El procedimiento requerido para generar habla artificial puede dividirse en dos etapas (King, 2011): la primera es convertir el texto en una especificación lingüística, ya que muchas representaciones en texto requieren una interpretación fonética para su generación. Por ejemplo, la lectura de números y símbolos se realiza interpretando primero estos símbolos antes de su pronunciación. El segundo proceso consiste en convertir esta especificación lingüística en la señal de audio con una clara correspondencia con el habla. Un sistema completo que enmarque ambas suele llamarse un sistema texto a habla (TTS).

El método dominante para realizar síntesis de voz durante las últimas décadas ha sido la selección de unidades (Suendermann, Höge y Black, 2010), en el cual se seleccionan segmentos de audio de longitud variable de una base de datos adecuadamente etiquetada y se unen para generar nuevas frases. Este método tiene requerimientos considerables de cantidad de datos almacenados y procesamiento necesarios. La síntesis de voz generada con técnicas recientes de aprendizaje profundo (Prenger y Catanzaro, 2019) también comparte esta característica de altos requerimientos de datos, en forma de grabaciones y textos, para realizar los procesos de ajuste en los complejos modelos matemáticos e informáticos en los que se basan.

Para muchas de las aplicaciones de esta tecnología es clara la conveniencia de generar voces con las cuales las personas usuarias puedan identificarse directamente. Por ejemplo, es importante considerar que un sistema de comunicación alternativo o aumentativo para una persona con discapacidad pueda producir habla en concordancia con la edad y el género de la persona usuaria. De forma semejante, asistentes virtuales y entornos educativos inteligentes pueden ser mejor aceptados por las personas usuarias y su entorno si éstos se comunican utilizando voces y acentos afines a quienes los utilizan. De esta manera surge la necesidad de crear los medios para la generación de habla artificial con múltiples acentos y que reflejan a todos los grupos etarios, incluyendo a niños y niñas.

En este trabajo se reportan los primeros resultados de generación de habla artificial con sonido de un niño y una niña en idioma castellano con acento costarricense, utilizando el método de síntesis estadística paramétrica de voz (Tokuda, Nankaku, Toda, Zen, Yamagishi, y Oura, 2013), surgido cerca del año 2000, y que se detalla en las siguientes secciones.

El resto del documento está organizado de la siguiente manera: En la sección 2 se esquematizan las generalidades del método de síntesis de voz utilizado. En la sección 3 se presentan los materiales y métodos utilizados. En la sección 4 se reportan los resultados obtenidos, y finalmente en la sección 5 se presentan las conclusiones.

## 2. Síntesis estadística paramétrica de voz.

Este tipo de síntesis, a diferencia de los modelos que manipulan directamente las señales de audio, se basa en la idea de representar las señales por un conjunto de parámetros, los cuales pueden ser aprendidos y luego reproducidos con modelos como los Modelos Ocultos de Markov (HMM), Este método contempla los siguientes pasos (Zen, Tokuda y Black, 2009):

- Convertir habla grabada en una representación paramétrica, que incluye coeficientes espectrales y de tono, así como aproximaciones de sus derivadas, y modelos de duración.
- Realizar un segmentación del audio en fonemas y un alineamiento de estos con los estados de un conjunto de HMM.
- Por cada HMM, entrenar un modelo estadístico del habla parametrizada, tomando en cuenta en el contexto del fonema.
- En la parte de síntesis, utilizar los HMM de forma generativa, para extraer la representación paramétrica de las nuevas frases que se desea sintetizar.
- Convertir en señal de audio la representación extraída.

El tipo de HMM que se utiliza en las implementaciones de esta técnica es el llamado "izquierda a derecha", pues cada uno de los estados del conjunto S tiene probabilidad de cambiar hacia sí mismo o hacia el estado de la derecha acorde con un conjunto de probabilidades $\pi$, como se ilustra en la Figura 1.

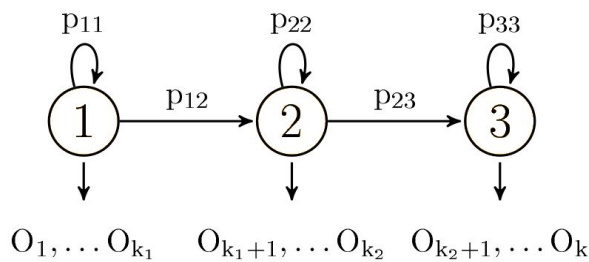

Figura 1: Ilustración de un HMM del tipo izquierda a derecha con tres estados.

De esta manera, los HMM pueden describirse mediante un vector $\lambda = (S, \pi_i, a, b)$, donde a es la matriz de probabilidades de transición entre estados y b la distribución de probabilidad de observaciones que se emitan al llegar a un estado. En este caso se asume que b sigue una distribución de probabilidad multivariada, definida como:

$$b_i(\boldsymbol{o}_t) = \frac{1}{\sqrt{(2\pi)^d |\boldsymbol{\Sigma}_i|}} \exp\left\{ \frac{-1}{2}(\boldsymbol{o}_t - \boldsymbol{\mu}_i)^\top \boldsymbol{\Sigma}_i^{-1}(\boldsymbol{o}_t) - \boldsymbol{\mu}_i \right\} \qquad (1)$$

donde $\mu_i$ y $\Sigma_i$ son vectores de medias y covarianza, respectivamente, d es la dimensión del vector de características y $o_t$ es el vector de observaciones que se emiten en el estado correspondiente, en el instante t.

De esta manera, en cada uno de los estados se emite un vector de características, tanto continuas (para la información espectral de la señal de habla), como multi-estado, llamadas así para describir la característica de la frecuencia fundamental en la voz, la cual cambia repentinamente desde valores cero hasta valores positivos, y luego de regreso a cero. El vector de características se ilustra en la Figura 2.

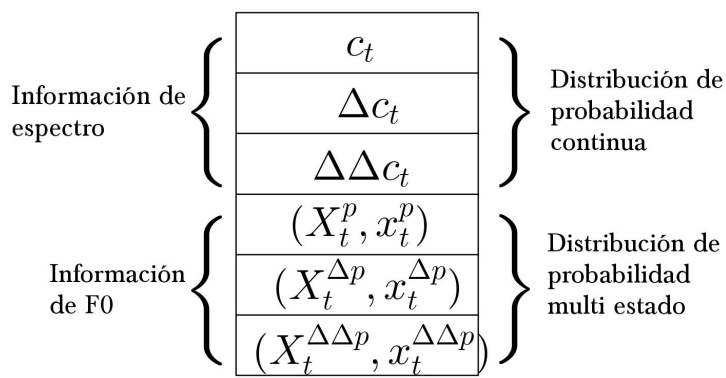

Figura 2: Vector de características emitido en cada estado de los HMM.

La síntesis estadística paramétrica tiene como ventaja su menor requerimiento de almacenamiento y procesamiento (Suendermann, Högey Black, 2010), así como su mayor flexibilidad. Esto debido a que al utilizar solamente la representación paramétrica, es decir, un modelo que consiste en conjunto de vectores y coeficientes, éstos pueden manipularse directamente.

La implementación de un sistema de síntesis de habla basada en esta técnica requiere un conjunto de programas, centrados en el software HTS (Tokuda, Zen, Yamagishi, Masuko, Sako, Black y Nose, 2008), así como de conocimiento de la especificación lingüística del idioma que se desea implementar. Los datos requeridos son un conjunto de audios y su correspondiente transcripción ortográfica, a partir de los cuales se generan los modelos, como se ilustra en la Figura 3.

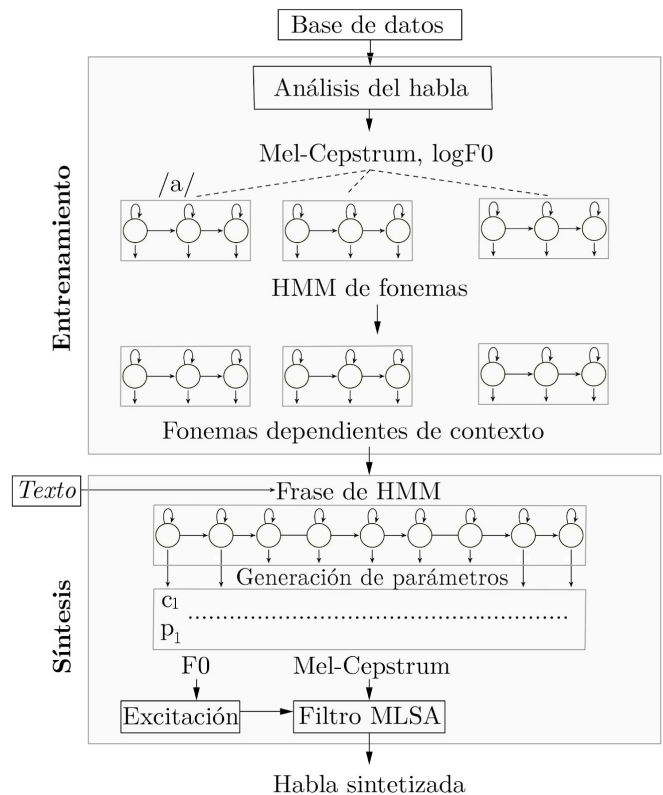

Figura 3: Esquema general de la síntesis basada en HMM

Si bien se han reportado recientemente en la literatura investigaciones de punta en síntesis de voz basadas en técnicas de aprendizaje profundo, la literatura de síntesis basada en HMM sigue siendo vigente, especialmente para idiomas y condiciones en las cuales no se cuentan con grandes cantidades de datos como para utilizar estos modelos, ni los basados en selección de unidades. Por ejemplo, en (Sefara, 2019) se reporta su utilización en sistemas con pocos recursos, mientras que en (Amrouche, Ahcéne y Falek, 2019; Reddy, Kiram y Sreenivasa, 2020; Zangar, 2020) se reporta su utilización en lenguas como el Hindi y el Árabe, donde aún no se exploran otras técnicas.

El caso de producción de habla artificial de niños con acento costarricense sin duda pertenece a los casos donde se cuenta con poca cantidad de información, por lo que la síntesis de voz basada en HMM se presenta como la opción viable para generar estos sonidos del habla y poder realizar una evaluación sobre ellos.

# 3. Materiales y métodos

### 3.1. Estrategia de recolección de datos

El proceso de recolección de datos se realizó en dos sesiones desarrolladas en los estudios de grabación del CEPROAV (Centro de Producción Audiovisual) de la Escuela de Ciencias de la Comunicación Colectiva de la Universidad de Costa Rica; en esta ocasión se contó con cinco participantes, dos niños con edades de seis y once años, así cómo tres niñas, dos de ellas con ocho años y una con edad de doce años.

Con respecto a las zonas de procedencia, dos de los masculinos y una de las femeninas eran oriundos de San

José, las otras dos femeninas eran de Alajuela.

Para esta grabación se contó con el consentimiento informado de las personas encargadas de los niños y niñas participantes, documento que fue elaborado teniendo como base el formato que dicta la Universidad de Costa Rica para este efecto.

Es importante destacar que este proceso de interacción fue de carácter exploratorio especialmente en atención a la población infantil, pues su participación en este tipo de estudios implican un reto especial principalmente por sus características particulares de desarrollo.

De igual manera, en el primer semestre del año 2019 se realizaron dos sesiones con jóvenes adultos estudiantes del nivel de bachillerato de las carreras de educación especial e ingeniería eléctrica. Entre las personas participantes se contó con tres masculinos con edades de diecisiete, dieciocho y veinte años, los dos primeros procedentes de Heredia y el tercero de Osa. También se contó con seis personas femeninas una de ellas con edad de veinte años, tres con edades de veintiún años, una con veintitrés años y una con veintiséis años.

### 3.2. Metodología de grabación

Con referencia al proceso de recolección de datos con las personas participantes menores de edad, se tomó en cuenta crear un ambiente de confianza y camaradería, abriendo el espacio para la comunicación libre y fluida relacionado con temas que les resultase de interés.

Se hizo uso de instrumentos formales y no formales de evaluación de la articulación, pero que para efectos del proyecto pretendían recopilar un compendio de voces humanas de personas de distintas edades y géneros.

Por lo tanto, se utilizó un test de articulación que permitió arrojar sonidos iniciales, medios y finales de cada uno de los fonemas que se encuentran en el alfabeto español. En este test se pudo observar distintas imágenes de fácil conocimiento y se le pedía a la persona menor participante que indicará su nombre, motivando así al uso de habla espontánea. Esta metodología de grabación es semejante a la descrita en (Jiménez y Rodríguez, 2020).

Dentro de las estrategias para la recolección de las voces infantiles fue necesario el uso de imágenes, tanto pictográficas como fotográficas, que pudiesen evocar en los y las menores la producción oral no solo del nombre de estas sino aspectos de agrupación semántica, el uso del género en las palabras pronunciadas así como si las mismas se encontraban en plural o singular.

Lo anterior aunado a los espacios de comunicación libre sobre temas de interés particular de cada persona menor participante, propició la recolección de una serie de datos consistentes que pudieron ser utilizados para el proceso de producción inicial de la voz artificial infantil.

### 3.3. Transcripción y segmentación

Posterior al proceso de grabación, se realizó un proceso manual de segmentación de los archivos de audio, para seleccionar las palabras adecuadas para aplicar las técnicas descritas en la Sección 2. Las palabras adecuadas fueron aquellas que contenían habla limpia, sin ruidos, palabras incompletas, risas u otros sonidos que quedaran registrados, de manera que la información auditiva tuviera una correspondencia clara con un texto de referencia.

Luego del proceso manual de segmentación, cada uno de los archivos resultantes de audio fue transcrito manualmente, de manera que se tuviera la correspondiente entre sonidos del habla y texto necesarios para el entrenamiento de los HMM. Tanto los archivos de audio como de texto son los insumos requeridos en el proceso de generación, descrito en la siguiente subsección.

### 3.4 Generación de voces

La base de datos, constituida por los conjuntos de archivos de audio y de texto descritos previamente para cada uno de los niños y niñas participantes, fue utilizada en conjunto con el sistema HTS (H.W. Group, 2013), en el cual están integrados los procesos de extracción de características y todos los algoritmos de entrenamiento de los modelos HMM.

El resultado de aplicación de este sistema es un conjunto de frases de habla artificial generada a partir de los parámetros de frecuencia fundamental y espectro en los estados de los HMM. Para realizar la evaluación, este conjunto de frases generadas fueron una réplica del contenido de las bases de datos, de manera que se tuviera un conjunto de elementos equivalentes para comparar ambos.

### 3.5. Evaluación

La evaluación de las frases resultantes consistió en valoraciones realizadas por un conjunto de 25 personas, las cuales escucharon audios correspondientes a las palabras generadas de forma artificial y las grabaciones de los niños, además de las mismas palabras aisladas pronunciadas por el grupo de adultos.

Se hizo una selección aleatoria de palabras de cada uno de estas fuentes, y se solicitó a las personas participantes, escuchar cada una y valorarla de acuerdo con los criterios:

1. Edad percibida en la voz: En dos categoría, voz infantil o voz adulta.
2. Género de la voz escuchada: En dos categorías, voz masculina o femenina.
3. Transcripción de la palabra escuchada: Una transcripción manual escrita de lo escuchado, para valorar su inteligibilidad.

## 4. Resultados

En el Cuadro 1 se muestran los resultados sobre la percepción de la edad percibido en los audios de habla. Se presenta en el orden en que fueron escuchados por las personas evaluadoras. En cada uno de los audios, estas personas podían seleccionar si escuchaban una voz de niño/niña, una voz de una persona adulta, o les parece que no se puede determinar a partir de la escucha.

Cuadro 1: Resultados de la identificación de edad en audios

| No. | Edad real | Cantidad de aciertos | Cantidad de errores | No se puede determinar | Tipo de voz |
|---|---|---|---|---|---|
| 1. | Infantil | 5 | 8 | 16 | Artificial |
| 2. | Infantil | 29 | 0 | 0 | Natural |
| 3. | Infantil | 29 | 0 | 0 | Natural |
| 4. | Infantil | 17 | 3 | 9 | Artificial |
| 5. | Infantil | 10 | 6 | 13 | Artificial |
| 6. | Adulta | 29 | 0 | 0 | Natural |
| 7. | Infantil | 24 | 4 | 1 | Natural |
| 8. | Infantil | 17 | 6 | 6 | Artificial |
| 9. | Adulta | 18 | 7 | 4 | Natural |
| 10. | Infantil | 29 | 0 | 0 | Natural |
| 11. | Infantil | 10 | 5 | 14 | Artificial |
| 12. | Adulta | 25 | 4 | 0 | Natural |
| 13. | Infantil | 14 | 1 | 14 | Artificial |
| 14. | Infantil | 14 | 12 | 3 | Natural |
| 15. | Infantil | 16 | 1 | 12 | Artificial |
| 16. | Adulta | 28 | 0 | 1 | Natural |

| 17. | Infantil | 29 | 0 | 0 | Natural |
|---|---|---|---|---|---|
| 18. | Adulta | 28 | 0 | 1 | Natural |
| 19. | Adulta | 29 | 0 | 0 | Natural |
| 20. | Adulta | 24 | 4 | 1 | Natural |

Fuente: Elaboración propia

Estos resultados se resumen en la Figura 4, en la cual se contabilizan y reportan como porcentaje de acuerdo con el tipo de voz. En esta Figura se puede ver cómo el porcentaje de aciertos presenta diferencias notables entre las voces artificiales y las naturales. Por su parte, también hay diferencias entre las voces naturales infantiles y las de adultos, de manera que se puede señalar con estos resultados que las personas pueden determinar con mayor claridad la pertenencia de una voz adulta a una voz infantil.

Si bien hay una proporción mucho mayor de escuchas que acertaron la voz natural que la infantil, el porcentaje de error en ambas es semejante. La mayor diferencia proviene de quienes indicaron que la edad en la voz artificial no se puede determinar.

Figura 4: Resultados de la evaluación realizada por la escucha de las frases de habla artificial y natural.

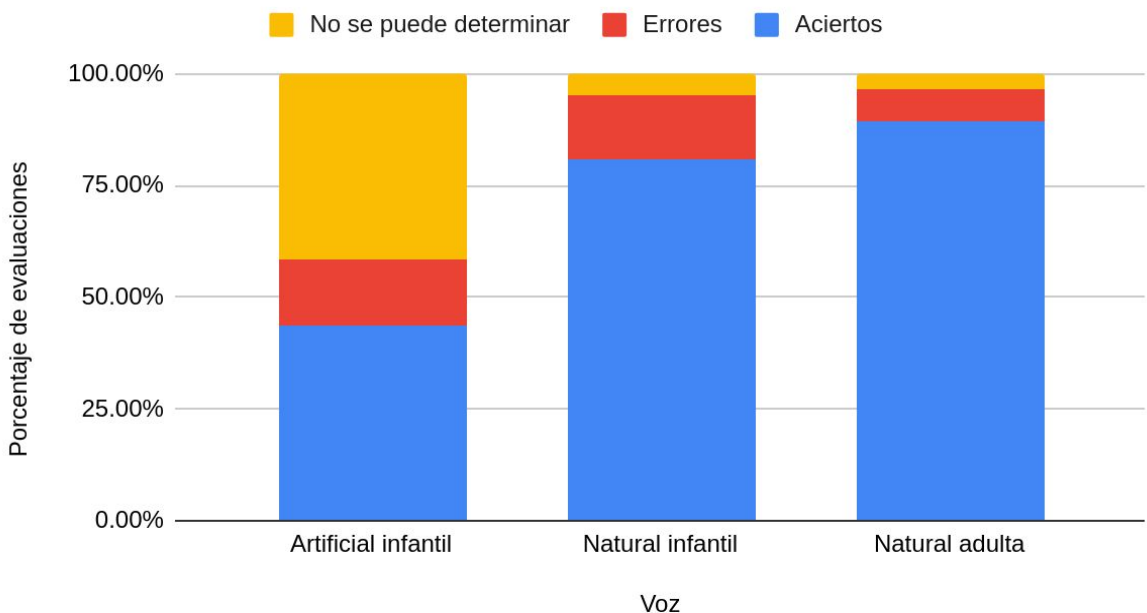

Fuente: Elaboración propia

En el Cuadro 2 se muestran los resultados sobre la percepción del género del hablante percibido en los audios de habla. Como en el caso anterior, se presenta en el orden en que fueron escuchados por las personas evaluadoras. En cada uno de los audios, estas personas podían seleccionar si escuchaban una voz masculina, una voz femenina, o que no se puede determinar el género a partir de la escucha.

Cuadro 2: Resultados de la identificación de género en audios

| No. | Género | Cantidad de aciertos | Cantidad de errores | No se puede determinar | Tipo de voz |
|---|---|---|---|---|---|
| 1. | Masculino | 2 | 17 | 10 | Artificial |
| 2. | Femenino | 11 | 11 | 7 | Natural |
| 3. | Masculino | 27 | 0 | 2 | Natural |
| 4. | Femenino | 6 | 6 | 17 | Artificial |
| 5. | Masculino | 2 | 6 | 21 | Artificial |
| 6. | Masculino | 29 | 0 | 0 | Natural |
| 7. | Femenino | 11 | 15 | 3 | Natural |
| 8. | Femenino | 8 | 6 | 15 | Artificial |
| 9. | Femenino | 26 | 2 | 1 | Natural |
| 10. | Femenino | 23 | 1 | 5 | Natural |
| 11. | Masculino | 3 | 12 | 14 | Artificial |
| 12. | Femenino | 27 | 2 | 0 | Natural |
| 13. | Femenino | 8 | 0 | 21 | Artificial |
| 14. | Femenino | 14 | 11 | 4 | Natural |
| 15. | Femenino | 4 | 3 | 22 | Artificial |
| 16. | Masculino | 29 | 0 | 0 | Natural |
| 17. | Masculino | 23 | 4 | 2 | Natural |
| 18. | Masculino | 29 | 0 | 0 | Natural |
| 19. | Masculino | 29 | 0 | 0 | Natural |
| 20. | Femenino | 28 | 0 | 1 | Natural |

Fuente: Elaboración propia

Estos resultados se resumen en la Figura 5, en la cual se contabilizan y reportan como porcentaje de acuerdo con el tipo de voz. En esta Figura se puede ver cómo el porcentaje de aciertos presenta diferencias aún mayores que el caso de la estimación de edad, entre las voces artificiales y las naturales. En este caso la proporción de errores también es más significativa para el caso de voces artificiales que en las naturales.

Figura 5: Resultados de la evaluación realizada sobre el género del hablante, en las frases de habla artificial y natural.

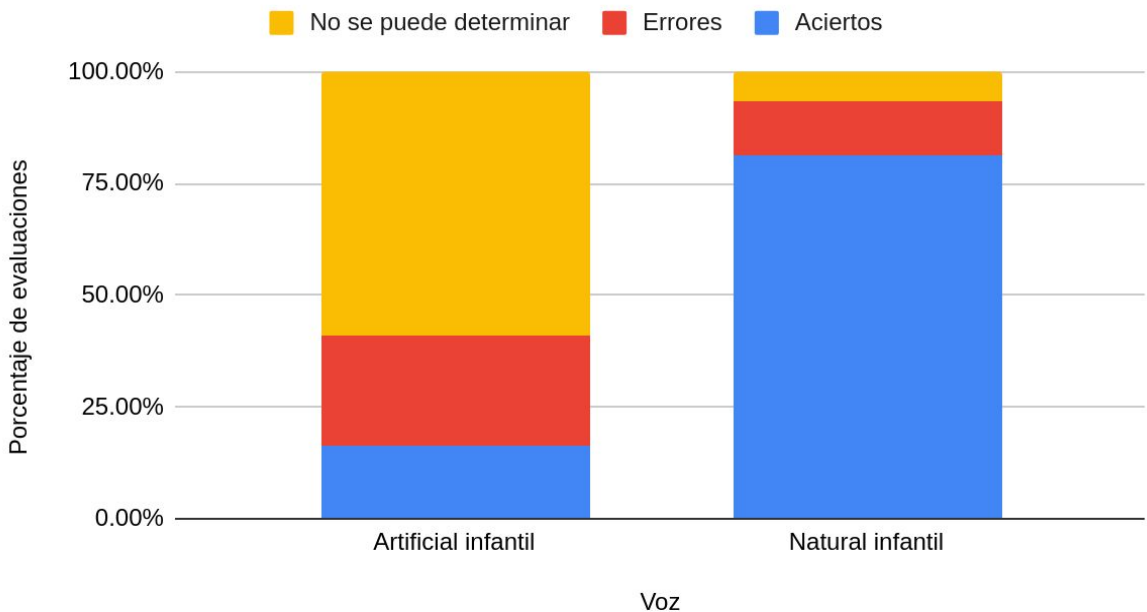

Fuente: Elaboración propia

Finalmente, en el Cuadro 3 se muestran los resultados de las transcripciones realizadas por las personas evaluadoras. En este caso, estas transcripciones solamente pueden evaluarse como correctas o incorrectas. Aquellos casos en que no se respondió a la pregunta, se interpretó como que no se entendió la palabra y se contabiliza como error.

Cuadro 3: Resultados del proceso de transcripciones de audios, por parte de las personas evaluadoras

| No. | Palabra | Cantidad de aciertos | Cantidad de errores | Tipo de voz |
|---|---|---|---|---|
| 1. | Clavo | 14 | 6 | Artificial |
| 2. | Pala | 20 | 0 | Natural |
| 3. | Tenis | 10 | 10 | Artificial |
| 4. | Cuello | 20 | 0 | Natural |
| 5. | Tenis | 9 | 11 | Artificial |
| 6. | Escoba | 20 | 0 | Artificial |
| 7. | Basura | 20 | 0 | Artificial |
| 8. | Dedos | 20 | 0 | Natural |
| 9. | Nariz | 12 | 8 | Artificial |
| 10. | Diente | 18 | 2 | Natural |
| 11. | Globo | 12 | 8 | Artificial |
| 12. | Silla | 10 | 10 | Artificial |
| 13. | Pala | 20 | 0 | Natural |
| 14. | Gallina | 20 | 0 | Natural |
| 15. | Tortuga | 20 | 0 | Artificial |
| 16. | Cisne | 20 | 0 | Natural |
| 17. | Pantalón | 9 | 11 | Artificial |
| 18. | Nariz | 18 | 2 | Natural |
| 19. | Puerta | 20 | 0 | Artificial |
| 20. | Oveja | 19 | 1 | Natural |

Los resultados de las transcripciones se resumen en la Figura 6. En esta figura se puede observar cómo la inteligibilidad de los resultados del habla artificial es inferior a los del habla natural en un porcentaje cercano al 25%.

Figura 6: Resultados de las transcripciones realizadas sobre las frases de habla artificial y natural.

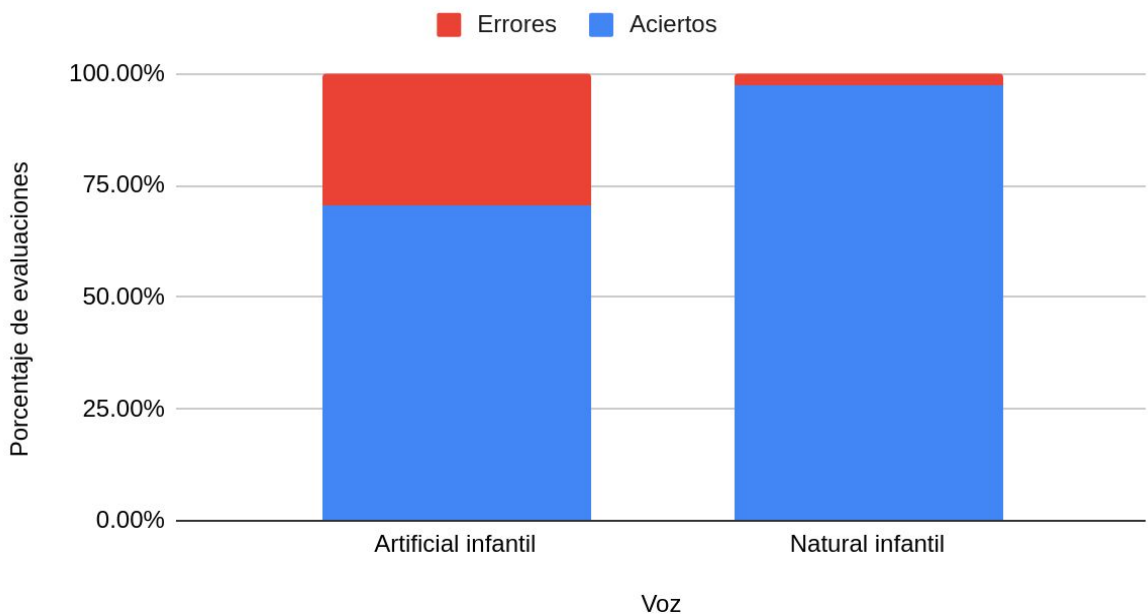

Fuente: Elaboración propia

# 5. Conclusiones

En el presente artículo se reporta la primera experiencia de generación de voces artificiales de niños y niñas con acento costarricense, utilizando técnicas estadísticas paramétricas basadas en HMM. Dadas las dificultades del proceso, principalmente en la obtención de bases de datos extensas de grabación homogénea, el proceso de generar habla infantil presenta retos mayores a la generación de habla con sonido de personas adultas.

Los resultados obtenidos en esta primera experiencia, la cual tuvo como fuente de datos palabras aisladas grabadas por niños con edades entre 3 y 12 años, muestran que existen diferencias significativas entre la claridad con que se percibe el género y la edad de la voz, con respecto al habla natural de los niños. En mayor medida, el establecimiento del género presentó diferencias más marcadas que el caso de la edad.

De igual manera, las diferencias en inteligibilidad del habla son aún grandes con respecto a la inteligibilidad del habla natural. Todas estas valoraciones fueron realizadas por un grupo de personas adultas, quienes escucharon tanto muestras de habla artificial como natural, sin información de a cuál tipo correspondía cada escucha.

Estos resultados constituyen una primera referencia de la capacidad de las técnicas estadísticas paramétricas para generar voces artificiales con las condiciones presentadas, y en este grupo etáreo en particular. Como trabajo futuro, se debe procurar la grabación de conjuntos mayores de datos y la aplicación de técnicas de mejora del sonido generado, los cuales pueden someterse a nuevos procesos de valoración subjetiva y comparar los avances con los presentadas en este estudio.

# Referencias